# Doped graphene oxide functionalization strategy for synthesis of nanocomposite membranes: electrospun coatings in biomedical field application

*Luz M. Rivera-Rivera[a, b], Netzahualpille Hernández-Navarro[c], Lina M. Hoyos-Palacio[a], Romeo de Coss[d], Nancy E. Ornelas-Soto[c]\*, Alejandra García-García[b]\*.*

[a] Grupo de Investigación en Biología de Sistemas, Escuela de Ciencias de la Salud, Universidad Pontificia Bolivariana, Circular 1a # 70-01, Medellín CP 05004, Colombia.
[b] Parque PIIT, Laboratorio de Síntesis y Modificación de Nanoestructuras y Materiales Bidimensionales, Centro de Investigación en Materiales Avanzados S.C., Alianza Norte #202, Carretera Monterrey-Aeropuerto Km 10, Apodaca, Nuevo León CP 66628, México.
[c] Centro del Agua para América Latina y el Caribe, Tecnológico de Monterrey, Campus Monterrey, Av. Eugenio Garza Sada 2501 Col. Tecnológico, Monterrey, NL CP 64849, México.
[d] Departamento de Física Aplicada, Centro de Investigación y estudios Avanzados del IPN, Km. 6 Antigua carretera a Progreso Apdo. Postal 73, Cordemex, 97310, Mérida, Yuc., México

*Corresponding authors: *ornel@itesm.mx* and *alejandra.garcia@cimav.edu.mx*

**ABSTRACT**

An electrospun membrane for vascular stent coating made of 2-methacryloyloxyethyl phosphorylcholine and butyl methacrylate copolymer (MPC-co-BMA), also known as (PMB), reinforced with functionalized and nitrogen doped reduced graphene oxide (f-NrGO) is presented. This membrane due to the nitrogen doped reduced graphene oxide (NrGO) negative electric charge has the capacity to repeal low density lipoproteins (LDL), which under specific conditions are the main cause of atherosclerotic disease. The NrGO functionalization process is detailed, as it creates strong bonds between NrGO and PMB, avoiding NrGO sheets to detach from the membrane. The copolymer synthesis was characterized by FTIR and chemical bonds between f-NrGO and PMB were proved by XPS and HNMR. Additionally, a simplified test bank that simulates blood flow conditions demonstrated the NrGO functionalization effect over the membrane. For the electrospinning process, optimal parameters were a voltage of 14.5 kV, and a flow rate of 0.3 mL/h, which lead to better properties of the membrane for the application. DMA results confirmed that the best reinforcement percentage of f-NrGO in terms of mechanical properties was 0.1 wt% and AFM images indicated the presence of the f-NrGO sheets over the fibers. Finally, the contact angle revealed the repulsion response to LDL, such behavior is promising to applications like cardiovascular coated stents.

## 1. INTRODUCTION

Cardiovascular disease is the leading cause of death and morbidity worldwide, and in most cases, the underlying cause of the cardiovascular event is atherosclerosis [1]. This disease is associated with elevated levels of low density lipoproteins (LDL), which are negative charged molecules that carry the cholesterol through blood [2]. Some risk factors for atherosclerosis are common conditions as dyslipoproteinaemia, diabetes, smoking, hypertension, and genetic abnormalities [3], leading to blood vessel wall (endothelium) damage, increased production of LDL and LDL oxidation. The formation of the atheroma plaque is triggered by the tendency of the oxidized LDL to penetrate the endothelium developing a complex process entailing a chronic inflammation of the arterial wall [4].

Currently, a well stablished treatment for atherosclerosis is the implantation of stents, small tubular devices that mechanically reopen the obstructed artery [5]. Nevertheless, these devices do not prevent the LDL to get into contact and penetrate the endothelium, this is the reason why the stent by itself cannot stop the progression of the atherosclerosis condition, making this treatment only reactive but not preventive. Some of these stents are drug-eluting devices or have polymer covers [6], but they are mainly focused on controlled drug delivery to promote anti-mitotic, antiproliferative, and anti-inflammatory responses [7]. For these reasons, new approaches are needed for improving stent implantation results, avoiding LDL to get into contact with blood vessel walls.

For the fabrication of different cardiovascular medical devices, polymers with specific characteristics that effectively prevent undesirable biological responses are needed, some of these desired characteristics are biocompatibility, hemocompatibility, antithrombogenicity and suppression of protein adsorption and cell adhesion [8]. It is well known that 2-methacryloyloxyethyl phosphorylcholine (MPC) polymers reach these standards, accordingly, these polymers have been used for several years to modify the surfaces and improve the overall biocompatibility in medical devices [9]. Copolymerization of MPC with other vinyl compounds by conventional [10] or advanced living radical polymerization [11] have been studied. MPC easily copolymerizes with styrene and with various alkyl methacrylates such as n butyl methacrylate (BMA), this copolymer known as PMB is considered one of the most important copolymers in this family [12].

PMB is highly used in biomedical applications, most of the times as coating for metal, ceramic or plastic substrates, whereby is suitable for the development of a reinforced stent coating membrane, processed by specific techniques such as electrospinning, which is a very used technique due to the simplicity and inexpensive nature of the setup needed [13], offering fiber diameters on the order of a few micrometers down to the tens of nanometers [14]. The capacity to easily produce materials at this biological size scale, control of pore structure, and obtention of high surface area fibers, have created a renewed interest in electrospinning for biomedical applications [15].

As well as polymer science, advances in carbon related materials are also protagonists in this field. These advances have driven the development of many interdisciplinary investigations. Within these materials is graphene, a single-layer carbon sheet with a hexagonal packed lattice structure [16], which has emerged as a 2D material with a large number of attributes that make it suitable even for medical applications. Its unique nature with exceptional mechanical, thermal, electronic properties and its ability to allow modifications in its structure, in addition to its biocompatibility, make it an ideal material to solve some of the most recurring medical problems [17].

However, as large amounts of graphene cannot be easily obtained [18], other 2D materials of graphene family may be used, as reduced graphene oxide, which can be produced by the reduction of graphene oxide, eliminating most of the oxygen functional groups [19], causing fascination in the scientific community, for the simplicity of the synthesis process, and other advantages as diverse doping and functionalization possibilities [20]. Moreover, pi-pi bonds present in graphene related materials, which are perpendicular to the lattice [21], can lead to interesting electronic properties such as electron conduction and a negative charged surface due to the formation of an electronic cloud. For all these reasons graphene, GO and rGO emerge as alternatives to modify existing materials and medical devices and give them new properties and applications.

This work presents the steps to follow for the nitrogen doped graphene oxide functionalization with cysteamine (2-aminoethanethiol), and the electrospinning parameters to generate a highly negative surface membrane, suitable for LDL repulsion. Obtained membranes could be proposed as future stent coating.

## 2. MATERIALS AND METHODS

### 2.1 NrGO synthesis

NrGO was obtained by chemical exfoliation of graphite powder from Merck without purification, following the methodology reported previously [22], using sulfuric acid, nitric acid, and potassium permanganate in constant stirring for 24 hours, followed by a process of simultaneous nitrogen doping and thermal reduction using ammonium hydroxide and GO in an autoclave at 200° C for 22 hours. This nitrogen doping confers to the material a highly negative electric charge surface capable of repealing negative charges from other materials or species.

### 2.2 NrGO functionalization

To generate strong bonds between NrGO and the copolymer, a functionalization process was carried out to incorporate cysteamine to the terminal functional groups of the NrGO. The cysteamine attaches from its amine end to the carboxylic groups of NrGO and from its thiol end to the copolymer. Hydrobromic acid 48% (HBr), oxalic acid, N-(3-Dimethylaminopropyl)-N-ethylcarbodiimide≥97%, phosphate buffered saline, *N*-Hydroxysuccinimide 98%, cysteamine hydrochloride ≥98% and ethyl alcohol were purchased from Merck and used without purification.

First, a stock of 30 mL of NrGO was prepared (2.5 mg/mL in aqueous solution), and sonicated for 1 h, then 5 mL of HBr were added under vigorous stirring at room temperature for 12 h, then, 1.50 g of oxalic acid were added and stirred for 4 h, the resulting material was filtered and dried at 50 °C under vacuum for 24 h. Thereafter 1 mg/mL of the resulting NrGO was dispersed in water, subsequently, 2.2 mmol/L of 1-ethyl-3- (3-dimethylamino) propyl carbodiimide (EDC) and 1.5 mmol/L of N-Hydroxysuccinimide (NHS) in a phosphate buffered saline (PBS) at a pH of 7.4 were added, finally the solution was treated with 130 µmol/L of cysteamine hydrochloride for 1 h and stirred at room temperature for 48 h. The resulting solution was purified by a 5 times centrifugation process at 14000 RPM to remove the saline solution and finally lyophilized, obtaining cysteamine functionalized NrGO (f-NrGO).

### 2.3 Copolymerization of PMB

For copolymerization process, 2-methacryloyloxyethyl phosphorylcholine (MPC), butyl methacrylate stabilized for synthesis (BMA), ethanol, azobisisobutyronitrile (AIBN) and petroleum ether were purchased from Merck and used without further purification.

1 g of MPC was added in a two-necked flask, immediately 1.237 g of BMA, 10 mL of ethanol and 18.8861 µL of AIBN as an initiator were added and the solution was stirred under a nitrogen atmosphere at 60 °C for 15.5 hours, after cooling, 200 mL of petroleum ether were added to separate the residues of the polymerization. Finally, the decanting was recovered and deposited in a Teflon™ paper inside a desiccator for 24 h (Figure 1).

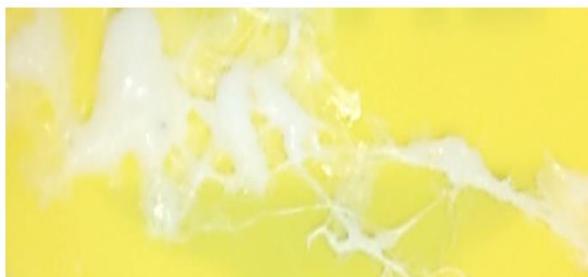

**Figure 1**. Physical aspect of dried PMB deposited on teflon paper.

### 2.4 Reinforcement of PMB with f-NrGO

Two different concentrations of f-NrGO powder (1 wt% and 0.1 wt%) relative to the amount of monomer were introduced into the copolymer following the same process previously described, but implementing a slight modification: f-NrGO powder was dissolved in the ethanol used for the polymerization process and this mixture was sonicated for 30 min. The other steps of the process continued without any modification. Figure 2 shows a photograph of the dry copolymer with 1 wt% f-NrGO, it is possible to appreciate a change in color from white (without f-NrGO in) to gray (with f-NrGO).

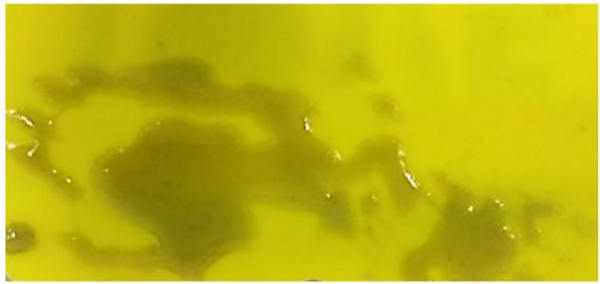

**Figure 2**. Physical aspect of dried f-NrGO reinforced PMB deposited on teflon paper.

### 2.5 Copolymer electrospinning process

A horizontal configuration electrospinning was used with a high voltage DC power supply (Gamma High Voltage Research Inc., USA) that reaches 30 kV and a syringe pump (New Era Pump System, Inc., USA) for the solution flow rate control; also, grounded 10 cm$^2$ aluminum plates were used as collectors.

Electrospinning of the polymer was carried out with a fixed distance from the needle to the collector (10 cm) and a fixed needle diameter (21G, external nominal diameter 0.8mm). All the fiber diameters mentioned in this work are the average of 150 fibers measured in 3 different samples. The fiber diameters were calculated using the free image processing software Image J.

A SEM analysis of the electrospun membranes (fiber morphology, fiber diameters and defects) was carried out in order to find the electrospinning conditions that generates a nonwoven membrane similar to biological tissue, such as highly interconnected fibers, presence of both, large and small fiber diameters (bimodal distribution), and few defects. These characteristics enhances biocompatibility, simulates native tissues, helps to reendothelialization and improves the physicochemical properties of the membrane [23], [24], [25], [26], [27]. Also, other studies report that the nanometric fibers (<100nm) improve the reendothelialization process [28], and promote controlled drug release [29]. Although a distribution with small and big diameters is desirable, fibers between 200 nm to 800 nm are needed in less quantity to ensure better mechanical resistance as suggested by DMA results, and to promote the formation of spider net behavior that improves biological performance [30].

First, PMB without f-NrGO was electrospun using different voltages (10.5 kV, 12.5 kV, 14.5 kV), flow rates (0.1 mL/h, 0.3 mL/h, 0.5 mL/h) and copolymer concentrations in ethanol (8 wt%, 10 wt%) to find out through the analysis of SEM micrographs, and membranes physical appearance the best conditions for electrospinning the material. A set of experiments was carried out first leaving the copolymer concentration fixed at 10 wt% and alternating the three different flow rates with the three different voltages, the sample with the best characteristics for the application was then analyzed with 8 wt% of copolymer concentration, finally PMB + 0.1 wt% f-NrGO and PMB + 1 wt% f-NrGO were electrospun using the previously identified electrospinning and fluid conditions to determine the f-NrGO concentration that promotes the formation a membrane similar to native tissue.

### 2.6 Low Density Lipoproteins oxidation

Low density lipoproteins from human plasma, copper (II) sulfate and ethylenediaminetetraacetic acid were obtained from Merck and used without further purification.

As LDL oxidation is promoted by the atherosclerosis disease, an artificial oxidation process is carried out to analyze the results under conditions similar to the real biological environment. Following the protocol reported by Chnari et al. [31], a high oxidation of the LDL was carried out as follows: 1 mL of LDL was mixed with 10 µg of $CuSO_4$, taken to a shaker in a water bath at 37 °C and the plate was covered with a glass urn to avoid heat loss; Then it was incubated for 18 hours and the process was finalized by adding a 0.01% w/v aqueous solution of EDTA to inhibit the oxidative activity on LDL.

### 2.7 Low Density Lipoproteins contact angle

Low density lipoproteins from human plasma were purchased from Merck. The wettability of each MPC specimen was evaluated by the measurement of the static contact angles of droplets of low density lipoproteins (5 µL each) on the sample surfaces. The contact angle measurements were carried out at five different locations of each specimen at room temperature. The results of the measurements are expressed as the mean of ten replicates.

### 2.8 Simplified test bank implementation for sheets detachment analysis

A simplified stent test bank was fabricated to experimentally validate the generation of chemical bonds between the f-NrGO sheets and the electrospun copolymer, also the membrane resistance to the simulated bloodstream flow was proved (Figure 3).

The test bank is a 1:1 prototype of a blood vessel. It uses simulated body fluid (SBF), at corporal temperature, this fluid is a solution with an ionic concentration similar to human plasma [32], and was prepared by dissolving the reagents: (NaCl, $NaHCO_3$, KCl, $K_2HPO_3·3H_2O$, $MgCl_2·6H_2O$, $CaCl_2$ and $Na_2SO_4$ in distilled water, the fluid was buffered to pH 7.4 at 36.5 °C with 50 mmol/dm$^3$ of $NH_2C(CH_2OH)_3$ and 45 mmol/dm$^3$ HCl 1M as reported previously [33]. Physiological conditions as aortic artery diameter and blood flow speed were reproduced during the tests using a 2 cm diameter glass tube and a water pump of 1300 L/h to keep the SFB flowing through the simulated vessel at the biological blood flow speed (30 cm/s to 60 cm/s [34]). Additionally, conditions were adjusted to ensure a laminar flow. Finally, a non-reactive chromium zirconium cooper wire was used to hold the membranes against the tube keeping them in position during the experiment.

Two types of membranes were analyzed, (PMB + functionalized NrGO electrospun membrane) and (PMB + non functionalized NrGO electrospun membrane). SBF was continuously flowing through the tube and in contact with each membrane for 24 hours, thereafter, a sample of SBF from each experiment was analyzed by Raman spectroscopy looking for the presence of NrGO sheets that may have detached from the membranes.

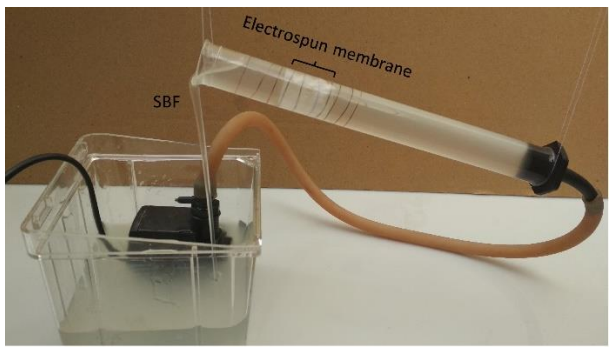

**Figure 3**. Simplified vascular test bank for sheets detachment analysis.

### 2.9 Characterizations

The morphology of the electrospun fibers was investigated by SEM (Fei Nova NanoSEM 200). The presence of the f-NrGO sheets over the fibers was confirmed by AFM (ASYLUM RESEARCH MFP3D-SA) in non-contact mode, with a speed range of 0.20 to 0.75 Hz and a silicone cantilever with a rectangular geometry of 240 mm in length, completely covered with a Ti/Ir film. A Thermo Scientific XPS spectrometer model Escalab 250Xi and Thermo Scientific iS50 infrared spectrometer were used for the identification of the NrGO changes produced by the functionalization process. For the determination of the thiol–polymer bonding H-NMR was performed in a Bruker spectrometer (300 MHz), spectra were acquired in 5 mm NMR tubes using deuterated methanol as solvent. Also, a Dynamic Mechanical Analysis was performed in a Q800 V21.3 Build 96 equipment, to investigate the mechanical properties of the membranes with a DMA module controlled force and 0.1 N/min, the samples had dimensions of 9.82 mm, 5.30 mm width and thickness respectively and 0.1 mm in length. The fluids from the test bank were analyzed using a high-resolution Raman model Labram HR from Horiba Jobin Yvon with a 632 nm laser. Finally, a Dataphysics OCA 15 Plus goniometer was used to measure the contact angle between the electrospun membranes and the oxidized LDL.

## 3. RESULTS AND DISCUSSION

### 3.1 NrGO functionalization

NrGO was functionalized with cysteamine to achieve strong chemical bonds between NrGO and the copolymer, this procedure aims to prevent the bloodstream from dragging the f-NrGO sheets out of the membrane once in contact with blood flow. The functionalization was carried out from the remaining carboxylic and epoxy functional groups of the NrGO [22]. First increasing the number of carboxyl groups, to finally adhere a linker molecule, cysteamine, with amine and thiol groups that can be anchored to both the NrGO and the copolymer.

NrGO functionalization mechanism is shown in Scheme 1. First, the remaining epoxy groups of the NrGO were transformed into hydroxyl groups through a ring opening reaction catalyzed by concentrated hydrobromic acid (Scheme 1a). The second step is the generation of an esterification reaction by introducing oxalic acid to interact with the hydroxyl groups reaching carboxylation (Scheme 1b). The third step is the activation of the carboxyl groups to increase the reactivity with the copolymer and the addition of the linker molecule (cysteamine), which has at one end amine groups that bind to the NrGO carboxyl groups by reacting with its OH part, turning into HN; the other end of the cysteamine molecule has thiol groups by which the molecule is anchored to the copolymer ending groups.

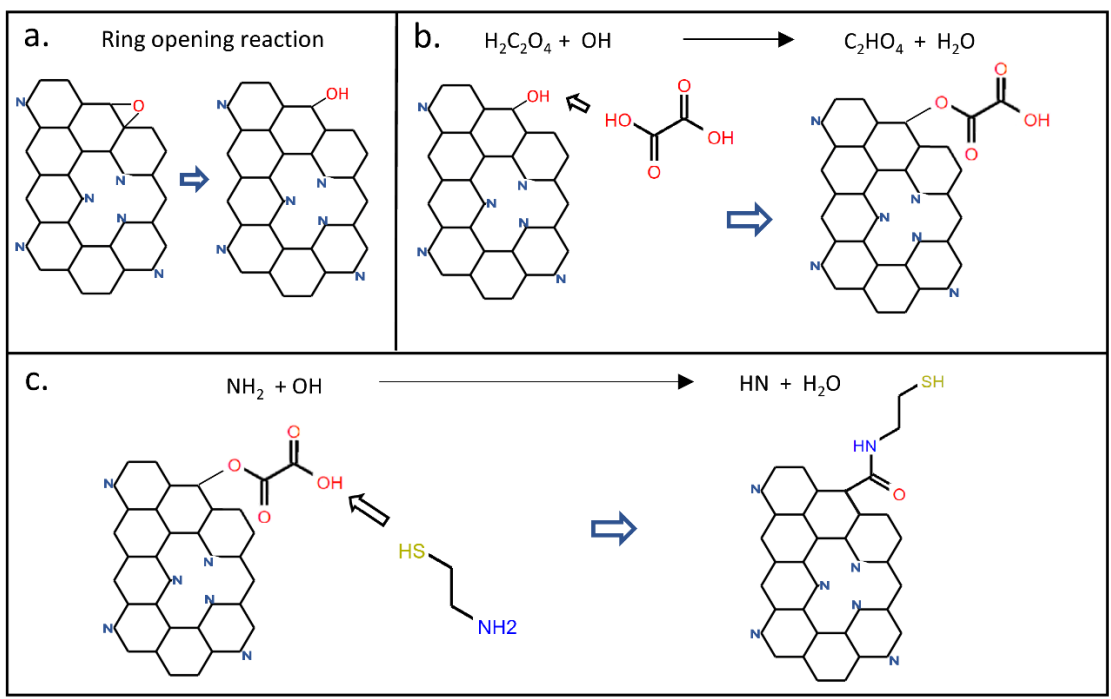

**Scheme 1.** NrGO functionalization with cysteamine. a) ring opening reaction – epoxy groups to hydroxyl groups b) carboxylation c) cysteamine reaction with NrGO carboxylic groups.

### 3.1.1 XPS characterization of the functionalized NrGO

XPS was used to analyze the first steps of the functionalization process. C1s and O1s core levels spectra of NrGO after carboxylation are presented in Figure 4a (deconvolution with a $\chi^2$ of 0.000295) and Figure 4b (deconvolution with a $\chi^2$ of 0.0003741) respectively, For both samples Shirley Sherwood and Slope background were used as well as Voigt curves (for symmetrical peaks) and Double Lorentzian curves (for asymmetric peaks). Figure 4a shows the presence of the contributions C=C $sp^2$ in 284.4 eV, the C-C $sp^3$ overlaps with the signal of C=N at 285.3 eV due to the proximity of these contributions that hinders the deconvolution of each one separately; The C-O functional groups appear at 286 eV, the C=O also overlaps with C-N at 287 eV, due to the proximity of each other. Finally, the contribution of the carboxyl groups appears at 290.4 eV [35], [36].

As the carboxylation step of the functionalization process aims to increase the number of carboxylic groups in the sheets, a comparison of the areas of the peaks corresponding to these groups before [22] and after carboxylation was performed. Table 1 shows these areas for C1s and O1s core levels before and after carboxylation. For C1s the NrGO treated with oxalic acid increases the peak area in 1179.49, similarly the O1s deconvolution (Figure 4b), also shows an increase of 4655.2 in the COOH peak area after the addition of oxalic acid, indicating that this step of the functionalization process generates more carboxylic groups on the NrGO surface, enhancing the possibilities of attachment with the cysteamine molecule in the final step of the functionalization.

**Table 1.** Areas of COOH contribution in XPS for C1s and O1s core levels of NrGO before and after carboxylation

| Core level | COOH peak area before carboxylation | COOH peak area after carboxylation |
|---|---|---|
| C1s | 1500.31 [22] | 2955.45 [22] |
| O1s | 2679.8 | 7610.65 |

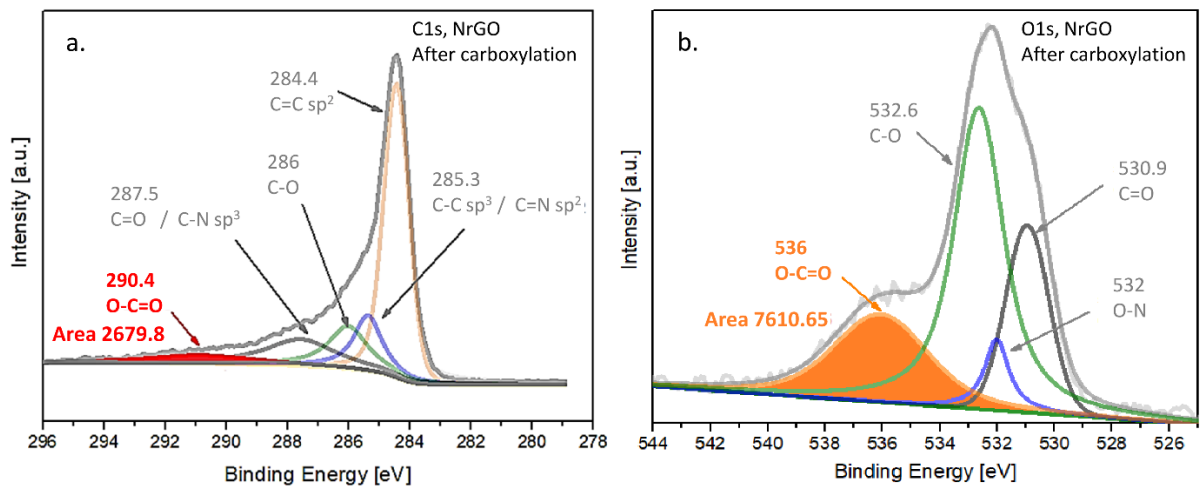

**Figure 4.** High resolution XPS spectra: deconvoluted peaks for the a) C1s and b) O1s corresponding to NrGO after carboxylation.

### 3.1.2 FTIR characterization of the functionalized NrGO

A characterization via FTIR-ATR was performed as shown in Figure 5, where the NrGO and the functionalized NrGO spectra are compared. The contribution at 2400 cm$^{-1}$ is attributed to the stretching vibrations of the SH bonds [37] present in the linker molecule cysteamine, this contribution appears only in the functionalized sample proving the presence of the cysteamine. Moreover, the peak that contains the contributions of C-N and C=N is widened only for the functionalized NrGO, this behavior is attributed to the presence of the NH bonds at 1557 cm$^{-1}$. Also, the intensity diminution of the band corresponding to OH at 3300 cm$^{-1}$ after the functionalization, is attributed to cysteamine bonding with the carboxylic groups that turn its OH into NH, giving a strong indication of the sample changes due to the functionalization process.

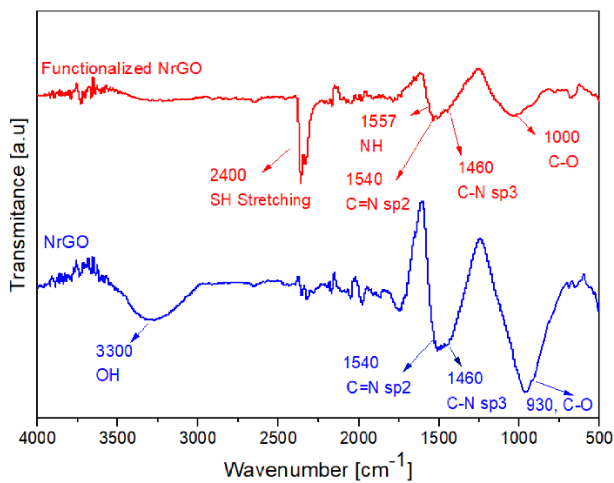

**Figure 5.** FTIR of NrGO and cysteamine functionalized NrGO, samples were dry to avoid water interference.

### 3.2 PMB characterization

MPC was copolymerized with BMA to improve the elasticity capacity [38]. To confirm this copolymerization the material was characterized by FTIR-ATR (Figure 6). The spectrum shows a broad peak detected around 3380 cm$^{-1}$ attributed to the OH bonds vibrations of the hydroxyl groups present in the polymer structure. The intense peaks at 2934 cm$^{-1}$, 2853 cm$^{-1}$, and the peak at 1474 cm$^{-1}$ are attributed to the stretching vibrations of the $CH_2$ groups of both MPC and BMA, the contribution of the BMA $CH_3$ asymmetric stretching and $CH_3$ symmetric stretching appears at 2956 cm$^{-1}$ and 2869 cm$^{-1}$ respectively. Additionally, the presence of a peak at 1721 cm$^{-1}$ corresponds to the stretching vibration of the $-(C=O)=-O-$ of the MPC and BMA esters, the three peaks at 1235 cm$^{-1}$, 1154 cm$^{-1}$ and 1065 cm$^{-1}$ are attributed to phosphorylcholine $-POCH_2-$. Inoue, et al. [39], have reported similar wavenumbers for these phosphorylcholine characteristic peaks (1240 cm$^{-1}$, 1160 cm$^{-1}$, and 1080 cm$^{-1)}$. Also, the same peaks were found in another study [28] with slight differences in the wavenumbers: OH peak was found at 3400 cm$^{-1}$, the stretching vibrations of the $CH_2$ were found at 2959, 2866, and 1474, the stretching vibration of the phosphorylcholine esters was found at 1724 cm$^{-1}$ and the $-POCH_2-$ vibrations were located at 1234, 1152, and 1076 cm$^{-1}$. Moreover, the peak at 1371 cm$^{-1}$ that corresponds to the $CH_3$ bending of butyl methacrylate, it is in agreement with other published studies [38,40]. In this way, it was verified that the material obtained is the copolymer PMB.

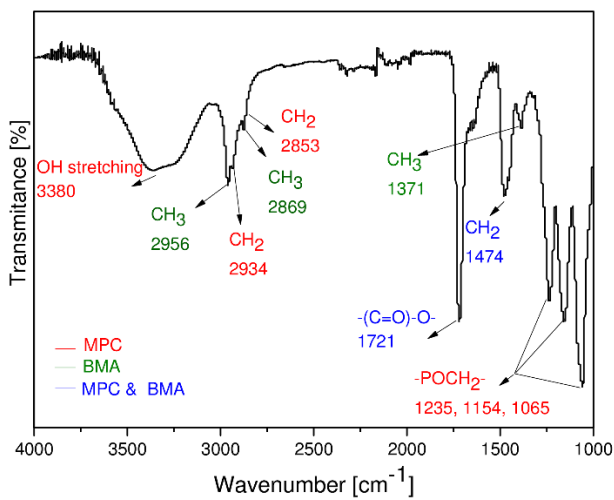

**Figure 6**. FTIR spectrum of copolymerized PMB *(MPC-co-BMA)*.

### 3.3 H-NMR of PMB and f-NrGO reinforced PMB

The bond between NrGO terminal groups and the amine end of the cysteamine molecule was verified by XPS, nevertheless, it is also important to prove that the other end of the cysteamine molecule (the thiol end) is chemically bonded to the copolymer, which will lead to a strong bond between NrGO and the PMB, ensuring good adhesion characteristics for the application, because the permanence of the NrGO over the polymer is crucial to avoid the NrGO detachment from the polymeric membrane.

H-NMR was conducted on both samples, PMB with f-NrGO (Figure 7a) and PMB without f-NrGO (Figure 7b) to verify the attached groups. For both samples the intense singlet at 3.30 ppm corresponds to the deuterated methanol; this singlet appears widened, behavior that can be attributed to the presence of the polymer amine that also appears at 3.30 ppm, for both samples the doublets at 1.9 ppm correspond to the $CH_3$ of the butyl methacrylate, and the contributions between 1.4 and 2.1 ppm are attributed to phosphorylcholine and butyl methacrylate $CH_2$. The multiplet between 3.5 ppm and 3.8 ppm corresponds to $CH_2$ in the central part of the copolymer structure and the multiplet between 3.9 ppm and 4.4 ppm is attributed to the phosphorylcholine $OCH_2$ $CH_2$ OP [40].

Nevertheless, the most important contributions to corroborate the formation of a chemical bond between the PMB and the cysteamine functionalized NrGO are the double-doublets that appear in 5.7 ppm and 6.25 ppm corresponding to the geminal hydrogens (two hydrogens linked to the same carbon), attributed to the double bond between C and terminal $CH_2$ of PMB Figure 7a). These contributions are present in PMB spectrum but disappear in PMB + f-NrGO spectrum indicating that the double bond between C and $CH_2$ is no longer there due to the reaction between the copolymer and the cysteamine thiol group (present in f-NrGO). These geminal hydrogens usually appear with small coupling constants (j) with values between 0 Hz and 3 Hz, this constant is the separation between any two peaks of each of two coupled multiple signals [41], in this case, the geminal hydrogens in the PMB appear with coupling constant of 1.46 Hz between peaks a1 - a2 and a coupling constant of 1.68 Hz between peaks b1 - b2 (left inset in Figure 7a), these values were calculated with equation 1. Where J is the coupling constant, x2 corresponds to the ppm of one of the peaks of the second doublet, x1 corresponds to the ppm of one of the peaks of the first doublet and η is the strength of the NMR spectroscope in MHz.

$$J = (x2 \cdot \eta) - (x1 \cdot \eta) \quad (1)$$

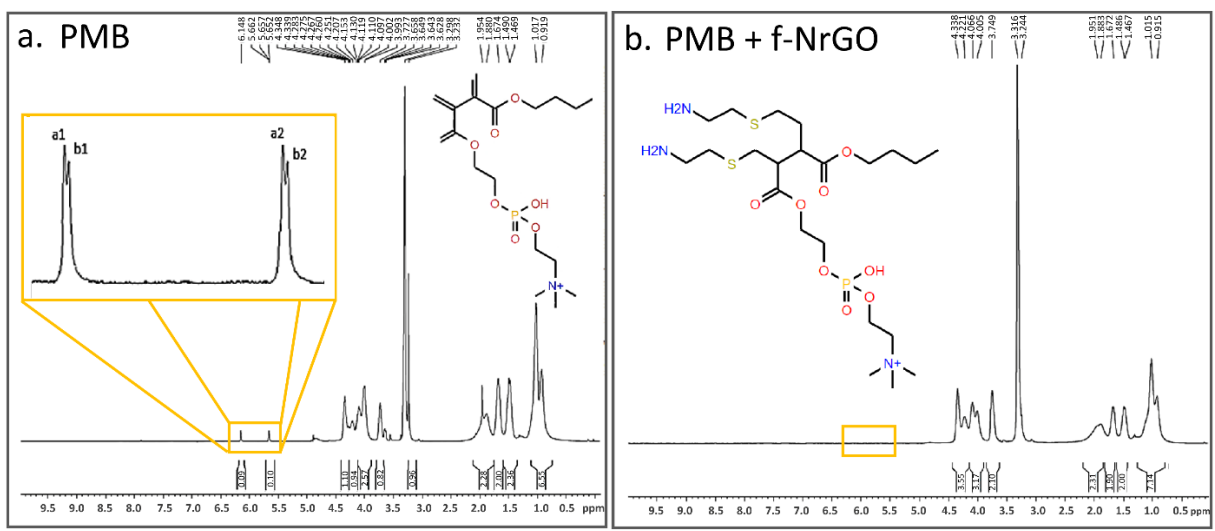

**Figure 7.** H-NMR a) PMB, left inset shows the geminal hydrogens and right inset shows the PMB structure b) f-NrGO reinforced PMB, left inset shows PMB structure bonded with cysteamine.

### 3.4 Obtention of the f-NrGO reinforced PMB membrane by electrospinning

Figure 8a shows the physical appearance of the best electrospun PMB sample, which was obtained with a voltage of 14.5 kV, a flow rate of 0.3 mL/h, and a copolymer concentration in solvent of 10 wt%. Figure 8b presents the SEM micrograph of the same sample revealing the presence of some fiber junctions with few morphology defects (beads) and Figure 8c shows a wide fiber diameter distribution for electrospun PMB where diameters from 60nm to 900 nm can be observed. The 55% of the fiber diameters are distributed in the ranges of 60 nm – 150 nm and 800 nm - 900 nm. Fibers with middle diameters (200 nm to 800 nm) are present but in less quantity, thus reaching the desired bimodal fiber diameter distribution, according to the proposed application.

In addition to the PMB electrospinning, PMB with two different f-NrGO concentrations (0.1 wt% and 1 wt%) were electrospun using the best previously identified voltage, flow rate, and copolymer concentration. It is important to highlight that f-NrGO incorporation during the polymerization achieves not just a physical but a chemical bond with the PMB as proved by H-NMR results, due to the possibility of interaction between the thiol group of cysteamine molecules and terminal groups of PMB, this chemical bond is fundamental to prevent f-NrGO agglomeration, as other studies propose [23], because proper incorporation of sheets through electrospun nanofibers by simple blending with polymer solution prior to electrospinning is a very difficult task due to sheets size and stacking of them. The formation of a copolymer / f-NrGO chemical bond showed great importance because it avoids the agglomeration of sheets during electrospinning.

Both electrospun samples, PMB + 1 wt% f-NrGO and PMB + 0.1 wt% f-NrGO present a very homogenous physical appearance as shown in Figure 8d and Figure 8g. SEM micrographs (Figure 8e and Figure 8h) show crossover fibers with the presence of junctions and wide bimodal distribution of diameters that is possible to appreciate in Figure 8f and Figure 8i, all these are desired characteristics for the application, leading to the simulation of native tissues [23]. Nevertheless, the PMB + 1wt% f-NrGO sample has a range of diameter variation from 134 nm to 1000 nm but most of the fibers have diameters nearer to 1000 nm (44% of the fibers are bigger than 800 nm and only 15% are smaller than 200 nm). The lack of more fibers within or near the nanometric range could be a problem due to the diminution of the reendothelization process, this fiber diameter distribution is attributed to the high f-NrGO reinforcement percentage, that may lead to a bad mixing with the copolymer creating agglomeration points that changes the fluid characteristics and negatively impact the electrospinning results. On the other hand, the sample with 0.1 wt% f-NrGO has the widest range of fiber diameter variation, ranging from 60 nm to 1350 nm. Fibers with diameters from 60 nm to 200 nm represent the 30% and fibers between 800 nm to 1350 nm the 43%, this feature gives the membrane all the benefits the nanometric fiber diameters but also allows the formation of spider net like junctions and improves the mechanical properties.

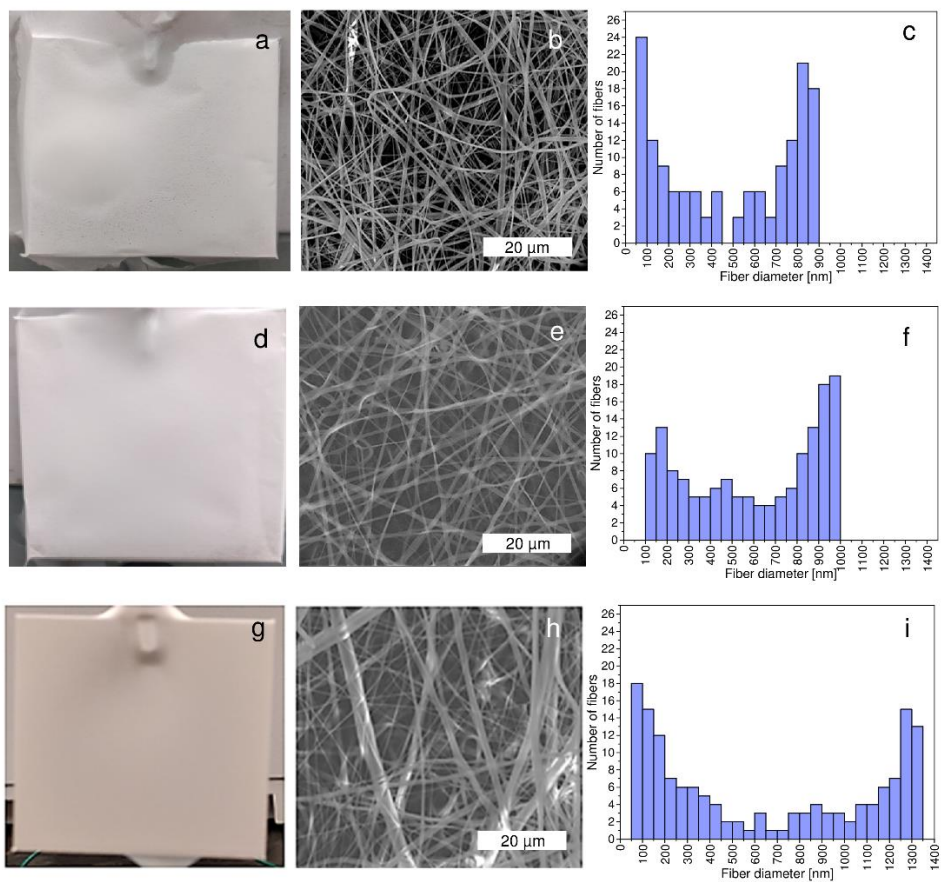

**Figure 8**. (14.5 kV, 0.3 mL/h, 10 wt% PMB concentration) a) Physical aspect of electrospun PMB. b) SEM micrograph of electrospun PMB. c) fiber diameter distribution of electrospun PMB. d) Physical aspect of electrospun PMB + 1wt% f-NrGO. e) SEM micrograph of electrospun PMB + 1wt% f-NrGO. f) fiber diameter distribution of electrospun PMB + 1wt% f-NrGO. g) Physical aspect of electrospun PMB + 0.1wt% f-NrGO. h) SEM micrograph of electrospun PMB + 0.1wt% f-NrGO. i) fiber diameter distribution of electrospun PMB + 0.1wt% f-NrGO.

### 3.5 f-NrGO reinforced PMB membrane characterizations

Differential mechanical analysis (DMA) tests were performed to analyze the mechanical behavior of electrospun PMB reinforced with 1 wt% and 0.1 wt% f-NrGO, then the presence and the distribution of the f-NrGO sheets over the PMB electrospun fibers were determined by AFM, also a contact angle test was carried out to analyze the effect of the f-NrGO reinforcement on the electric charges repulsion and finally, the membrane was proved with a test bank to experimentally verify the effect of the functionalization in sheets capacity to remain attached to the copolymer while in contact with a fluid.

### 3.5.1 Differential mechanical analysis for f-NrGO reinforced PMB membranes

DMA analysis was carried out to determine the stress-strain curve of three samples: PMB, PMB + 0.1 wt% f-NrGO and PMB + 1 wt% f-NrGO as shown in Figure 9, where changes in mechanical properties generated by PMB reinforcement with different percentages of f-NrGO are evident.

All samples exhibit a typical polymer mechanical behavior, the PMB and PMB + 0.1 wt% f-NrGO showed a characteristic behavior of flexible plastics (plastic, non-linear) and the PMB + 1 wt% f-NrGO behaves as a rigid plastic (linear and without plastic zone). The main parameter to observe in these graphs is the Young Modulus that expresses the stiffness of a material, where a larger value means that the material is more rigid [42], other important parameters are stress and strain at fracture, these help to verify the effect of the copolymer reinforcement with f-NrGO and its influence on the ability of the coating to withstand stress efforts, due to expansion-contraction conditions to which a stent is subjected [43].

In the stress-strain curve of the PMB and the PMB + 0.1 wt% f-NrGO there is a non-linear plastic behavior, however, both have a tendency to linearity in the proportionality zone, for this type of material, it is possible to calculate the Young modulus through the use of various methods such as obtaining the slope of a secant line, tangent method, and chord method, in addition to linear approximations with respect to the curve in the proportionality zone [24]. To obtain the Young modulus in the non-linear samples, the average result of all the previously mentioned methods was calculated. For the sample with linear behavior (PMB + 1 wt% f-NrGO), the Young modulus was determined from the slope of the elastic zone. The obtained Young modulus values for the three samples are presented in Table 2 where it is possible to see that the maximum Young modulus value (24.5 MPa) is held by the PMB with the maximum amount of f-NrGO (1 wt%). Before reinforcement, the electrospun PMB had a Young Modulus of 15.2 MPa, which means that a high percentage of f-NrGO turns the material more rigid, this is consistent with the behavior presented in the stress strain curve where the reinforcement with 1 wt% f-NrGO drastically changes the mechanical response of the material turning it into a rigid plastic. Once the elastic zone ends the material reaches fracture, and this happens at a lower elongation than unreinforced PMB (2.23% and 9.5% elongation respectively). These results suggest that the reinforcement with 1 wt% f-NrGO decreases the resistance by stress efforts, this behavior is attributed to the increase in the f-NrGO sheets percentage causing a non-homogeneous distribution in the copolymer during its polymerization due to the increased viscosity. On the other hand, when comparing the unreinforced PMB and the PMB reinforced with 0.1 wt% f-NrGO, it is possible to observe in Table 2 that the Young Modulus decreases from 15.2 MPa to 9.55 MPa, this shows that the addition of f-NrGO in moderate amounts provides flexibility to the material making it more suitable for the desired application, both values are within the reported range of young modulus accepted for stent covers (0.5 MPa to 20 MPa) [44], but the addition of 0.1wt% f-NrGO to the PMB membrane considerably lowers the Young modulus making this membrane more suitable to be adjusted to many kinds of stents. Additionally, it increases the fracture strength as well as the fracture strain, which goes from

9.50% to 33.6%. Therefore, the PMB reinforced with 0.1 wt% f-N-rGO has more convenient mechanical properties to withstand the expansion-contraction stress during the stent positioning process.

These results are in agreement with other studies [23] which show that polymers with low GO related materials reinforcement achieved improved mechanical properties of membranes [30]. Nevertheless, studies as the presented by Hwang and Yu [45] conclude that lower diameter electrospun fibers achieve higher Young´s modulus due to the negligible shear deformation. In the present work, the electrospun sample (0.1 wt% f-N-rGO) has small diameters but it still has a low Young´s modulus. This may be explained by the presence not just small and homogeneous diameters, but a wide distribution (big and small diameters) in the same sample. This will lead to the simulation of native tissues and not only to improve mechanical properties.

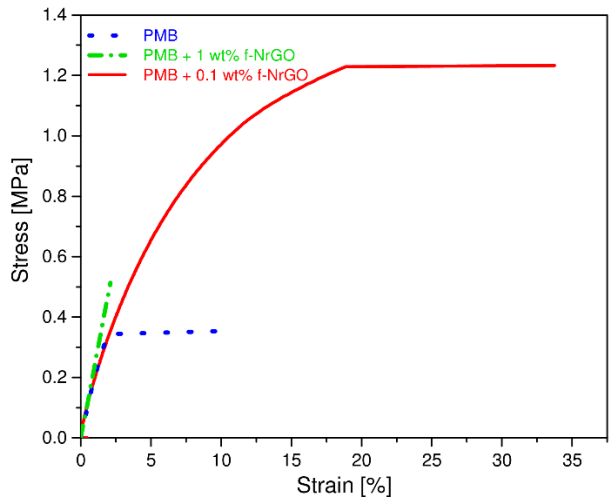

**Figure 9.** DMA of PMB, PMB + 1wt% f-NrGO and PMB + 0.1wt% f-NrGO electrospun samples.

**Table 2.** Average Young's Modulus for PMB, PMB + 1wt% f-NrGO and PMB + 0.1wt% f-NrGO electrospun samples.

| **Electrospun Sample** | **Average Young Modulus (MPa)** |
|---|---|
| PMB | 15.2 |
| PMB + 1 wt% f-NrGO | 24.5 |
| PMB + 0.1 wt% f-NrGO | 9.55 |

### 3.5.2 Atomic force microscopy of the electrospun membranes

To verify the presence and distribution of the f-NrGO sheets on the electrospun fibers, AFM characterizations were performed. Figure 10 shows the electrospun PMB reinforced with 0.1% f-NrGO. The protrusions over the fibers are attributed to the deposition of f-NrGO sheets distributed on the surface of the copolymer, this was verified by scanning Kelvin probe microscopy, finding different surface potentials for the protrusions than for the rest of the fibers, thus corroborating that the electrospinning process did not destroy the sheets and that these are adhered to the fiber. This distribution is attributed to the selected process for incorporating f-NrGO sheets to the PMB during the polymerization process and not just a mechanical mixing with the solvent once the copolymer is ready. This procedure avoids agglomeration suffered by the GO related materials due to the chemical bonding with the PMB chains, which leads to a homogeneous distribution of the sheets in the copolymer.

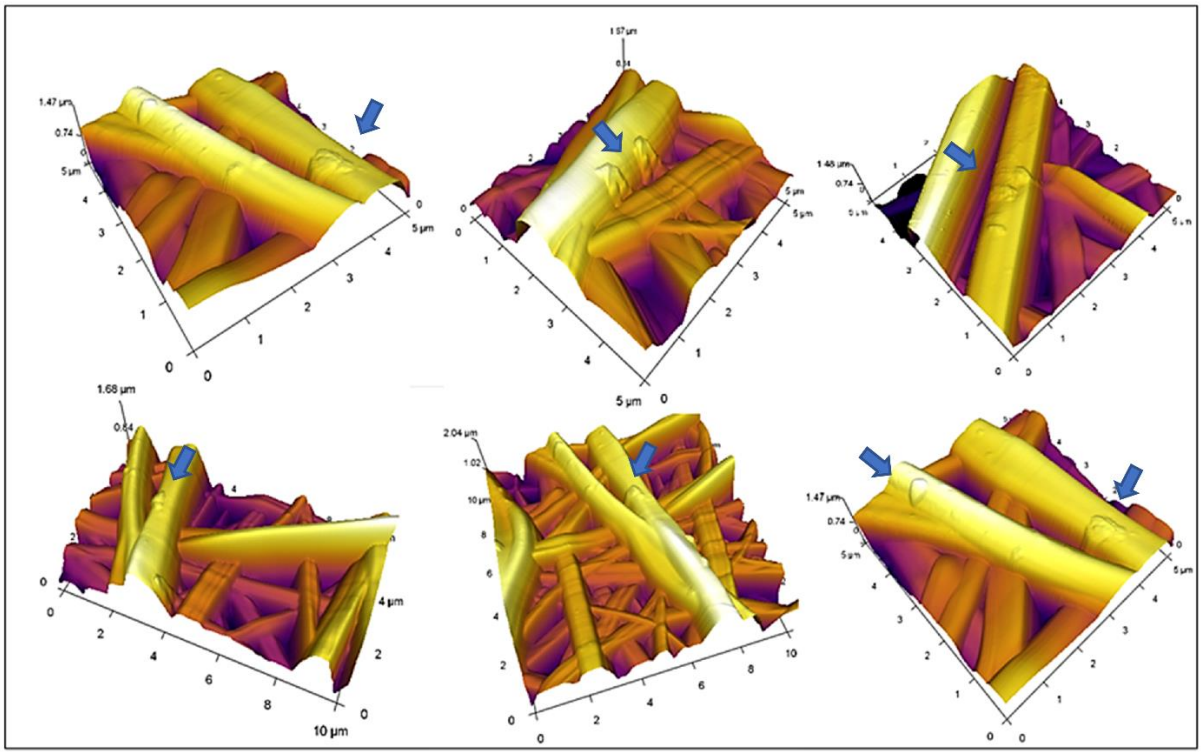

**Figure 10**. AFM images of the electrospun sample with 0.1 wt% functionalized-NrGO

### 3.5.3 Electrospun PMB and PMB + 0.1wt% f-NrGO LDL Contact Angle

Figure 11 presents the photographs of the best contact angle results of oxidized LDL upon electrospun PMB (Figure 11a), and oxidized LDL on electrospun PMB + 0.1 wt% f-NrGO (Figure 11b). The first one, with a 104.1° contact angle, corresponding to a lipophobic behavior attributed to the phosphorylcholine nature, nevertheless, the second one has a much greater angle (142.7°).

This contact angle almost reaches superlipophobicity and it is attributed to the presence of f-NrGO sheets over the copolymer fibers, enhancing coating capacity to repel negative electric charges, due to the nitrogen present in the f-NrGO lattice generating a negative charged electronic cloud at the membrane surface. The contact angle tests were performed over three different membranes with the same characteristics using 10 drops per sample; obtaining an average contact angle for the oxidized LDL upon electrospun PMB of 103.4° and an average contact angle for oxidized LDL on electrospun PMB + 0.1 wt% f-NrGO of 142.3°. These results demonstrate the membrane capacity to avoid accumulation of LDL in highly oxidized states, thus helping to prevent the progression of the atherosclerotic disease.

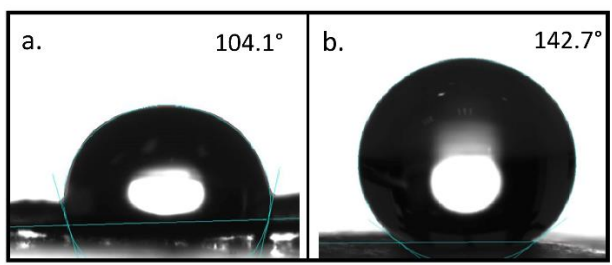

**Figure 11.** Contact Angle a) electrospun PMB with oxidized low density lipoproteins, b) electrospun PMB + 0.1 wt% f-NrGO with oxidized low density lipoproteins.

### 3.5.4 Raman Spectroscopy Validation of the Functionalization Process

A simplified test bank was used to test the sheets attachment in 2 different membranes, PMB + f-NrGO and PMB + NrGO (the last sample was obtained following the exact same process explained in the methodology section but using NrGO without cysteamine functionalization). The SBF in contact with these membranes was analyzed by Raman spectroscopy. Figure 12a, Figure 12b and Figure 12c show the Raman spectrum of the SBF before starting the experiments, the SBF that was in contact with the f-NrGO reinforced PMB membrane, and the SBF that was in contact with the NrGO reinforced PMB membrane respectively. It is notable that the Raman spectrum, of the fluid before the experiment and the fluid in contact with the f-NrGO reinforced membrane is very similar, presenting the typical contributions of the elements present in the SBF which simulates the blood plasma [46]. These contributions appear between 500 cm$^{-1}$ and 1800 cm$^{-1}$ however, the spectrum of SBF in contact with non-functionalized NrGO reinforced membrane shows predominant contributions at 1327 cm$^{-1}$ and 1599 cm$^{-1}$ corresponding to D and G bands of NrGO. Additionally, for this sample a widened peak in 2700 cm$^{-1}$ can be observed, also corresponding to the 2D band of NrGO [47]. Other peaks are present in the range of 500 cm$^{-1}$ to 1000 cm$^{-1}$ as well as 1700 cm$^{-1}$ to 1800 cm$^{-1}$ which are attributed to the SBF, and the SBF peaks that should be between 100 cm$^{-1}$ and 1700 cm$^{-1}$, are overlapped by the D and G bands of NrGO. These results suggest that the spectrum of the SBF in contact with non-functionalized NrGO PMB membrane corresponds to a combination of SBF and NrGO, thus verifying that the non-functionalized NrGO sheets did not adhere correctly to the PMB and detaches from the membrane when it comes into contact with the SBF during the experiment. On the other hand, the membrane reinforced with the functionalized NrGO resisted the test without NrGO sheets detaching from the copolymer, as the Raman spectrum only shows presence of SBF. These results strengthen the XPS and HNMR findings that shows the chemical attachment between NrGO and PMB through cysteamine functionalization and suggest that the sheets will not detach from the polymer fibers under specific conditions such as contact with fluids.

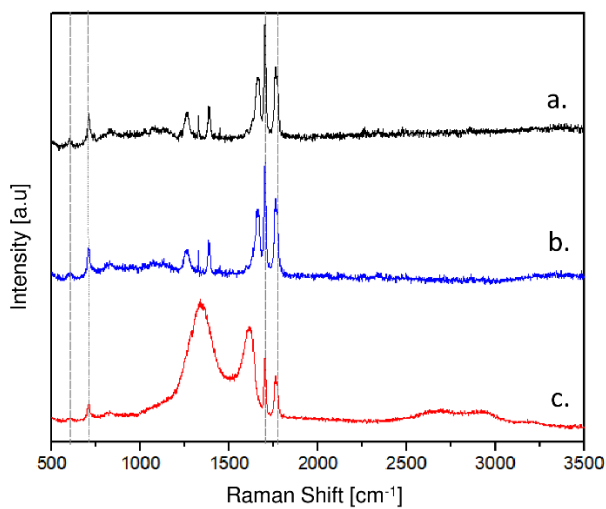

**Figure 12.** Raman spectrum a) Simulated body fluid b) Simulated body fluid in contact with electrospun f-NrGO reinforced PMB c) Simulated body fluid in contact with electrospun NrGO reinforced PMB.

## 4. CONCLUSIONS

Cysteamine functionalization of NrGO ensures chemical bonds between NrGO and PMB due to the reaction of its amine end with NrGO carboxylic groups, and its thiol end with PMB terminal groups, thus avoiding sheets detachment from the membrane. This functionalization process also contributes to the distribution of the sheets in the electrospun copolymer, leading to a good distribution of them over the fibers during electrospinning. It is also notable that the percentage of f-NrGO reinforcement in the copolymer plays a crucial role in membrane mechanical properties; 0.1 wt% f-NrGO reinforcement lowers electrospun PMB Young modulus and enhances its stress and strain at fracture, behavior attributed to the synergy between the fiber diameter distribution and good distribution of sheets in the copolymer given by the right amount of bonds between f-NrGO and PMB, which are enough to ensure a strong attachment but not so much to mitigate the elongation capacity of the polymer. This reinforcement percentage is also enough to generate a high repulsion to negative electric charges as the ones present in LDL, due to the highly negative surface of the NrGO distributed all over the electrospun fibers, characteristic that is retained even after the functionalization process. According to these results the present membrane is promising for its application as cardiovascular stent coating.

**INTERESTS STATEMENT**

Authors have no competing interests to declare.

**AUTHOR CONTRIBUTIONS**

The manuscript was written through contributions of all authors. All authors have given approval to the final version of the manuscript.

**ACKNOWLEDGEMENTS**

To CONACYT and COLCIENCIAS for its financial support through the postgraduate scholarship. Authors thank to Nayely Pineda Aguilar, Luis Gerardo Silva Vidaurri, Alberto Toxqui Terán and Oscar E. Vega at Centro de Investigación en Materiales Avanzados S.C. and Winston Quiñones at Universidad de Antioquia for infrastructure and technical support. Also, thanks to G.I. Hernández-Bolio at Laboratorio Nacional de Nano y Biomateriales CINVESTAV- Unidad Mérida by NMR technical support.

**REFERENCES**

[1] Gisterå, A. Hansson, G. K. The immunology of atherosclerosis. Nat Rev Nephrol. 2017,13, 368–380.

[2] Mello, A. P. Q. da Silva, I. T., Abdalla, D. S. P. Damasceno, N. R. T. Electronegative low-density lipoprotein: Origin and impact on health and disease. Atherosclerosis. 2011, 215, 257-265.

[3] Bergheanu, S. C. Pathophysiology and treatment of atherosclerosis Current view and future perspective on lipoprotein modification treatment. Neth Heart J. 2017, 25, 231–242

[4] Weber, C. Noels, H. Atherosclerosis: current pathogenesis and therapeutic options. Nat. Med. 2011, 17, 1410-1422.

[5] Simard, T. et al. The Evolution of Coronary Stents: A Brief Review. Can. J. Cardiol. 2014, 30, 35-45.

[6] Thakkar, A. S. Dave, B. A. Revolution of drug-eluting coronary stents: an analysis of market leaders. EMJ. 2016, 1, 114-125

[7] Farhatnia, Y. Tan, A. Motiwala, A. Cousins, B. G. Seifalian, A. M. Evolution of covered stents in the contemporary era: clinical application, materials and manufacturing strategies using nanotechnology. Biotechnol. Adv. 2013, 31, 524-542.

[8] Ishihara, K. Blood compatible surfaces with phosphorylcholine- based polymers for cardiovascular medical devices. Langmuir. 2019, 35, 17778-1787.

[9] Ishihara, K. Revolutionary advances in 2-methacryloyloxyethyl phosphorylcholine polymers as biomaterials. J. Biomed. Mater. Res. Part A. 2019, 107, 933-943.

[10] Ueda, T. Oshida, H. Kurita, K. Ishihara, K. Nakabayashi, N. Preparation of 2-Methacryloyloxyethyl Phosphorylcholine Copolymers with Alkyl Methacrylates and Their Blood Compatibility. Polym. J. 1992, 24, 1259-1269.

[11] Ishihara, K. Tsuji, T. Kurosaki, T. Nakabayashi, N. Hemocompatibility on graft copolymers composed of poly (2-methacryloyloxyethyl phosphorylcholine) side chain and poly (n-butyl methacrylate) backbone. J. Biomed. Mater. Res. 1994 28, 225-232.

[12] Ishihara, K. Choi, J. Methacrylate Monomer, first ed. Boca Raton FL: Elsevier, 2010.

[13] Pham, Q. P. Sharma, U. Mikos, A. Electrospinning of Polymeric Nanofibers for Tissue Engineering Applications: A Review. Tissue Eng. 2006, 12, 1197-1211.

[14] Sill, T. J. & Recum, H. A. Von. Electrospinning: Applications in drug delivery and tissue engineering. Biomaterials 2008, 29, 1989-2006.

[15] Bhardwaj, N. Kundu, S. C. Electrospinning: A fascinating fiber fabrication technique. Biotechnol. Adv. 2010,28, 325-347.

[16] Huang, X. Qi, X. Zhang, H. Graphene-based composites. Chem Soc Rev. 2012, 41, 666-686.

[17] Rani, J. R. Oh, S. Jang, J. Raman Spectra of Luminescent Graphene Oxide (GO) -Phosphor Hybrid Nanoscrolls. Materials. 2015, 8, 8460–8466.

[18] Nolan, H. et al. Nitrogen doped reduced graphene oxide electrodes for electrochemical supercapacitors. Phys. Chem. Chem. Phys. 2014, 16, 2280.

[19] Alam, S. Sharma, N. Kumar, L. Synthesis of Graphene Oxide (GO) by Modified Hummers Method and Its Thermal Reduction to Obtain Reduced Graphene Oxide (rGO). Graphene. 2017, 6, 1–18.

[20] Chem, J. M. Reduced graphene oxide/nickel nanocomposites: facile synthesis, magnetic and catalytic properties. J. Mater. Chem. 2012, 22, 3471

[21] Skarker, S. Hexagonal Boron Nitrides (White Graphene): A Promising Method for Cancer Drug Delivery. Int. J. Nanomedicine. 2019, 14, 9983–9993.

[22] Rivera, L. et al. Simultaneous N doping and reduction of GO: Compositional, structural characterization and its effects in negative electrostatic charges repulsion. Diam. Relat. Mater. 2019, 97, 107447.

[23] Kim, J. Park, C. H. Kim, C. S. Processing and characterization of electrospun graphene oxide/polyurethane composite nanofibers for stent coating. Chem. Eng. J. 2015, 270, 336-342.

[24] Lord, J.D. Morrel, R. Measurement Good Practice Guide No. 98 Elastic Modulus Measurement. https://www.npl.co.uk/special-pages/guides/gpg98_elastic (accessed, April 9, 2020).

[25] Raj, H. et al. Photocatalytic TiO2–RGO/nylon-6 spider-wave-like nano-nets via electrospinning and hydrothermal treatment. J. Memb. Sci. 2013, 429, 225-234.

[26] Wang, X. Ding, B. Sun, G. Wang, M. Yu, J. Progress in Materials Science Electro-spinning / netting: A strategy for the fabrication of three-dimensional polymer nano-fiber / nets. Prog. Mater. Sci. 2013, 58, 1173-1243.

[27] Raj, H. Raj, D. Taek, K. Baek, W. Photocatalytic and antibacterial properties of a TiO2 / nylon-6 electrospun nanocomposite mat containing silver nanoparticles. J. Hazard. Mater. 2011, 189, 465-471.

[28] Beck, G. Crichton, H. J., Baer, E. Monte, M. Bioactive Surface Functionalization Preparation and Characterization of 2-Methacryloyloxyethyl Phosphorylcholine Polymer Nanofibers Prepared via Electrospinning for Biomedical Materials. J. Appl. Polym. Sci. 2014, 131, 40606

[29] Maeda, T. Hagiwara, K. Yoshida, S. Hasebe, T. Hotta, A. Preparation and Characterization of 2-Methacryloyloxyethyl Phosphorylcholine Polymer Nanofibers Prepared via Electrospinning for Biomedical Materials. J. Appl. Polym. Sci. 2014, 131, 40606: 1-6.

[30] Chen, Y. Qi, Y. Tai, Z. Yan, X. Zhu, F. Xue, Q. Preparation, mechanical properties and biocompatibility of graphene oxide / ultrahigh molecular weight polyethylene composites, Eur. Polym. J. 2012, 48, 1026–1033.

[31] Chnari, E. Lari, H. B. Tian, L. Uhrich, K. E. Moghe, P. V. Nanoscale anionic macromolecules for selective retention of low-density lipoproteins. Biomaterials. 2005, 26, 3749–3758.

[32] Oyane, A. Kim, H. Furuya, t. Kokubo, T. Miyazaki, T. Nakamura, T. Preparation and assessment of revised simulated body fluids. J Biomed Mater Res A. 2003, 65, 188-195.

[33] Cho, S. Nakanishi, K. Kokubo, T. Soga, N. Ohtsuki, C. Apatite Formation on Silica Gel in Simulated Body Fluid: Its Dependence on Structures of Silica Gels Prepared in Different Media. J. Biomed. Mater. Res 1996, 33, 145–151.

[34] Angelsen, B.A.J. Brubakk, A. O. Transcutaneous measurement of blood flow velocity in the human aorta. Cardiovasc. Res. 1976, 10, 368-379.

[35] Berendjchia, A. Khajavib, R. Yousefic, A. Yazdanshenasd, M. Surface characteristics of coated polyester fabric with reduced graphene oxide and polypyrrole. Appl. Surf. Sci. 2016, 367, 36–42.

[36] Bu, J. Huang, X Li, S. Jiang, P. Significantly enhancing the thermal oxidative stability while remaining the excellent electrical insulating property of low .... Carbon. 2016, 106, 218–227, 2016.

[37] Innocenzi, P. Infrared spectroscopy of sol – gel derived silica-based films: a spectra-microstructure overview. J Non-Cryst Solids. 2003, 316, 309-319.

[38] Li, Y. Yongkui, H. Guobin, D. Shuijin, Y. Catalytic Synthesis of N-butyl Methacrylate with H4SiW6Mo6O40/SiO2. Adv. Mater. Res. 2013, 632, 135-139.

[39] Inoue, Y. Onodera, Y. Ishihara, K. Biointerfaces Preparation of a thick polymer brush layer composed of poly ( 2-methacryloyloxyethyl phosphorylcholine ) by surface-initiated atom transfer radical polymerization and analysis of protein adsorption resistance," Colloid Surface B. 2016, 141, 507–512.

[40] Alothman, Z. A. Preparation and characterization of alkyl methacrylate capillary monolithic columns. J. Saudi Chem. Soc. 2014, 16, 271-278.

[41] D. Pavia, G. Lampman, G. Kriz, J. Vyvyan, Introduction to Spectroscopy, fourth ed. Washington, 2009.

[42] Breitmaier, E. In Structure Elucidation by NMR in Organic Chemistry. John Wiley & Son, West Sussex, 2003, Chapter 1, pp 1-10.

[43] Encyclopedia Britannica.; Young´s Modulus. https://www.britannica.com/science/Youngs-modulus (accessed, April 9, 2020).

[44] Tomberli, B. Mattesini, A. Balderschi, G. Di Mario, C. Breve historia de los stents coronarios A Brief History of Coronary Artery Stents," Rev. Esp. Cardiol. 2018, 71,312–319.

[45] Cully, E. Vonesh, M. Expandable covered stent with wide range of wrinkle-free deployed diameters," United States Patent. US 8.425,584 B2. 2013.

[46] Hwang, K. Y. Yu, W. R. Measuring Tensile Strength of Nanofibers using Conductive Substrates and Dynamic Mechanical Analyzer. Fiber Polym. 2009, 10, 703-708.

[47] Rekha, P. Aruna, P. Daniel, A. Prasanna, W. Udayakumar, K. Ganesan, S. Bharanidharan, G. Raman Spectroscopic Characterization of Blood Plasma of Oral Cancer. 2013 IEEE 4th International Conference on Photonics (ICP), Melaka, 2013, pp. 135-137.